# Quantum Rabi Oscillations in Coherent and Squeezed Coherent States Via One- and Two-Photon Transitions

Moorad Alexanian

*Department of Physics and Physical Oceanography*
*University of North Carolina Wilmington*
*Wilmington, NC 28403-5606*

**Abstract:** We determine the Rabi oscillations for coherent and squeezed coherent states via one- and two-photon atomic transitions in two- and three-level atoms, respectively. The effect of squeezing coherent states is to reduce the large number of photons of mesoscopic scale to fewer photons and so be able to use numerical solutions with the aid of the Jaynes-Cummings model. We maximize the squeezing of the coherent state by minimizing the ratio of the photon number variance to photon number, thus narrowing the region where revival occurs.

## 1. Introduction

The Jaynes-Cummings model (JCM) [1] of a two-level atomic system coupled to a single-mode radiation field is known to exhibit interesting optical phenomena, such as the collapse and revival of Rabi oscillations of the atomic coherence [2–5]. However, complex features may appear at the mesoscopic scale (few tens of photons) in the analytical analysis of the collapse and revival beyond the rotating wave approximation when the energy spectrum of the system is changed drastically [6–9]. Recently, collapse and revival has been observed in a range of interaction times and photon numbers using slow circular Rydberg atoms interacting with a superconducting microwave cavity [10]. This experiment opens promising perspectives for the rapid generation and manipulation of non-classical states in cavity and circuit quantum electrodynamics.

In this paper we consider numerically solutions of the JCM with the field originally in a coherent state as well as in a squeezed coherent state. The effect of squeezing a coherent state is to reduce the mean photon number from a mesoscopic scale (large number of photons in the cavity) to a microscopic scale (fewer photons in the cavity) that would be more in agreement with the numerical results from the JCM. In addition, we minimize the ratio of the photon variance to photon number to sharpen the regions where revivals occur.

This paper is arranged as follows. In Sec. 2, we use the Bogoliubov-Valatin transformation to generate the algebra of quasiparticles. In Sec. 3, we use the algebra of creation and annihilation operators of quasiparticles to generate coherent states of quasiparticles that are actually photon squeezed coherent states. In Sec. 4, the probability amplitude for photons in the quasiparticle coherent state are obtained. In Sec. 5, the quotient of the photon variance to the mean number of photons is minimized in order maximize the squeezing of the coherent state. In Sec. 6, collapse and revival for one- and two-photon transitions are obtained numerically showing the narrowing of the peaks of the Rabi oscillations owing to squeezing. In Sec. 7, we study the parity Rabi oscillations for both one- and two-photon transitions again showing the narrowing of the peaks owing to squeezing. Finally, Sec. 8 summarizes our results.

## 2. Bogoliubov-Valatin transformation

Consider the Bogoliubov-Valatin [11, 12] canonical transformation

$$\hat{A} = \beta \hat{a} + \gamma \hat{a}^\dagger \qquad \hat{A}^\dagger = \beta^* \hat{a}^\dagger + \gamma^* \hat{a}, \tag{1}$$

where $\hat{a}, \hat{a}^\dagger$ are the photon annihilation and creation operators, respectively, with vacuum state $\hat{a}|0\rangle = 0$. The creation and annihilation operators $\hat{A}$ and $\hat{A}^\dagger$ satisfy the commutation relation

$$\left[\hat{A}, \hat{A}^\dagger\right] = 1 \tag{2}$$

provided $|\beta|^2 - |\gamma|^2 = 1$.

The corresponding normalized vacuum state, viz., $\hat{A}|\mathbf{0}\rangle = 0$, is given by

$$|\mathbf{0}\rangle = \sqrt{1 - |\gamma/\beta|^2} \sum_{n=0}^{\infty} (-\gamma/\beta)^n \frac{\sqrt{(2n)!}}{2^n n!} |2n\rangle. \tag{3}$$

Therefore,

$$|\mathbf{0}\rangle = \hat{S}(\zeta)|0\rangle \tag{4}$$

where

$$\hat{S}(\zeta) = \exp\left(-\frac{\zeta}{2}\hat{a}^{\dagger 2} + \frac{\zeta^*}{2}\hat{a}^2\right) \tag{5}$$

is the squeezing operator with

$$\gamma/\beta = e^{i\varphi} \tanh(r) \tag{6}$$

and $\zeta = r\exp(i\varphi)$. Transformation (1) is not unitary and may be interpreted as introducing quasiparticles.

One can introduce a basis in the quasiparticle Hilbert space since the $n$ quasiparticle state given by

$$|\mathbf{n}\rangle \equiv \frac{1}{\sqrt{n!}}\left(\hat{A}^{\dagger}\right)^{n}|\mathbf{0}\rangle = \hat{S}(\zeta)|n\rangle, \quad (7)$$

with the aid of

$$\hat{S}(\zeta)\hat{a}\hat{S}(-\zeta) = \cosh(r)\hat{a} + e^{i\varphi}\sinh(r)\hat{a}^{\dagger} = \hat{A}, \quad (8)$$

where $|n\rangle$ is the $n$ photon Fock state. Therefore, the state $|\mathbf{n}\rangle$ represents $n$ squeezed photons and so the set $\{|\mathbf{n}\rangle\}$, $n = 0, 1, 2, ...,$ spans the whole Hilbert space.

### 3. Quasiparticle coherent state

One can generate a coherent state for the quasiparticles by the action on the vacuum $|\mathbf{0}\rangle$ of the Glauber displacement operator

$$\hat{\mathbf{D}}(\alpha)|\mathbf{0}\rangle = \exp\left(\alpha\hat{A}^{\dagger} - \alpha^{*}\hat{A}\right)|\mathbf{0}\rangle = \hat{D}(\alpha\cosh(r) - \alpha^{*}e^{i\varphi}\sinh(r))\hat{S}(\zeta)|0\rangle, \quad (9)$$

where

$$\hat{D}(\alpha) = \exp(\alpha\hat{a}^{\dagger} - \alpha^{*}\hat{a}) \quad (10)$$

with $\alpha = |\alpha|\exp(i\theta)$.

Now

$$\hat{D}(\alpha\cosh(r) - \alpha^{*}e^{i\varphi}\sinh(r))\hat{S}(\zeta) = \hat{S}(\zeta)\hat{D}(\alpha) \quad (11)$$

with the aid of

$$\hat{S}(-\zeta)\hat{a}\hat{S}(\zeta) = \cosh(r)\hat{a} - e^{i\varphi}\sinh(r)\hat{a}^{\dagger}, \quad (12)$$

and so

$$\hat{\mathbf{D}}(\alpha)|\mathbf{0}\rangle = \hat{S}(\zeta)\hat{D}(\alpha)|0\rangle. \quad (13)$$

Therefore, the coherent state of quasiparticles is the squeezed coherent state of photons. This correspondence is valid for any state, that is, a given state of quasiparticles is equal to the squeezed state of the same state albeit for photons.

## 4. Probability amplitude

The probability amplitude for photons in the quasiparticle coherent state follows from (13)

$$\langle n|\hat{S}(\zeta)\hat{D}(\alpha)|0\rangle = \frac{(\tanh(r)e^{i\varphi})^{n/2}}{2^{n/2}(n!\cosh(r))^{1/2}} \exp\left(-\frac{1}{2}\left(|\alpha|^2 - e^{-i\varphi}\alpha^2 \tanh(r)\right)\right) H_n\left(\frac{\alpha e^{-i\varphi/2}}{\sqrt{2\cosh(r)\sinh(r)}}\right), \quad (14)$$

where $H_n(z)$ is the Hermite polynomial [13].

The corresponding probability is given by

$$p(n) = \left|\langle n|\hat{S}(\zeta)\hat{D}(\alpha)|0\rangle\right|^2 = \frac{(\tanh(r))^n}{2^n n! \cosh(r)} \exp\left(-|\alpha|^2 + \frac{1}{2}\tanh(r)\left(e^{-i\varphi}\alpha^2 + e^{i\varphi}\alpha^{*2}\right)\right) \left|H_n\left(\frac{\alpha e^{-i\varphi/2}}{\sqrt{2\cosh(r)\sinh(r)}}\right)\right|^2. \quad (15)$$

## 5. Photon average and variance

The moments of the number operator $\hat{n}$ can be calculated with the aid of Mehler's formula [14]

$$\sum_{n=0}^{\infty} \frac{u^n}{2^n n!} H_n(x) H_n(y) = \frac{1}{\sqrt{1-u^2}} \exp\left(\frac{2uxy - u^2(x^2 + y^2)}{1-u^2}\right) \quad (16)$$

by differentiation with respect to $u$. Now,

$$\langle \hat{n}^l \rangle = \sum_{n=0}^{\infty} n^l p(n) \quad (17)$$

and so

$$\langle \hat{n} \rangle = \langle \hat{a}^\dagger \hat{a} \rangle = |\alpha|^2 e^{-2r} + \frac{1}{4}\left(e^r - e^{-r}\right)^2 \quad (18)$$

and

$$\Delta n^2 = \langle \hat{n}^2 \rangle - (\langle \hat{n} \rangle)^2 = |\alpha|^2 e^{-4r} + \frac{1}{8}\left(e^{2r} - e^{-2r}\right)^2, \quad (19)$$

where $p(n)$ is given by (15) and $\varphi = 2\theta$.

We minimize the ratio $\Delta n^2 / \langle \hat{n} \rangle$ when considering collapse and revival in order to obtain optimal squeezing of the coherent state. One obtains

$$|\alpha|^2 = \frac{1}{16}(e^{2r} - 1)\left(3 + 3e^{6r} + 3e^{4r} - 5e^{5r} - \sqrt{9e^{12r} + 18e^{10r} - 13e^{8r} - 28e^{6r} + 43e^{4r} - 14e^{2r} + 1}\right). \quad (20)$$

In the following section we consider the collapse and revival of the coherent state with $\langle\hat{n}\rangle = 24.6$ and the squeezed coherent state with $|\alpha| = 10.0$, which yields $r = 0.7136$ from (20) and $\langle\hat{n}\rangle = 24.6$ from (18). Fig. 1 shows the photon number probability distribution $p(n)$ for both cases. Note the effect of squeezing the coherent state whereby $\Delta n^2/\langle\hat{n}\rangle = 1$ for the coherent state while $\Delta n^2/\langle\hat{n}\rangle = 0.3125$ for the squeezed coherent state. Note also the reduction in the average photon number from $\langle\hat{n}\rangle = 100$ in the coherent state to $\langle\hat{n}\rangle = 24.6$ in the squeezed coherent state.

### 6. Collapse and revival

*a. One-photon transition*

The Jaynes-Cummings model [1] of a two-level atomic system coupled to a single-mode radiation field is known to exhibit interesting optical phenomena, such as the collapse and revival of Rabi oscillations of the atomic coherence [2–5]. For instance, if the atom is initially in the excited state $|2\rangle$ while the radiation field is in a superposition of photon numbers with probability amplitude $a_n$, then the probability for being in the ground state $|1\rangle$ at time $t$ is

$$p(t) = \frac{1}{2}\sum_{n=0}^{\infty}|a_n|^2\left[1 - \cos(2\lambda\sqrt{n+1}\,t)\right], \qquad (21)$$

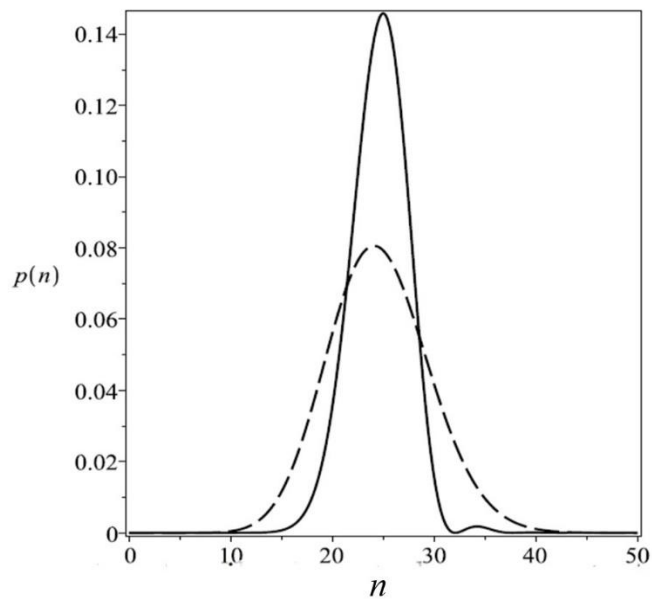

**FIG. 1:** Plot for the photon number probability distribution of the coherent state (dash) for $\langle\hat{n}\rangle = 24.6$ and the squeezed coherent state in Eq. (15) (solid) for $|\alpha| = 10$. Note that $\varphi = 2\theta$ and (20) gives $r = 0.7136$ and (18) gives $\langle\hat{n}\rangle = 24.6$.

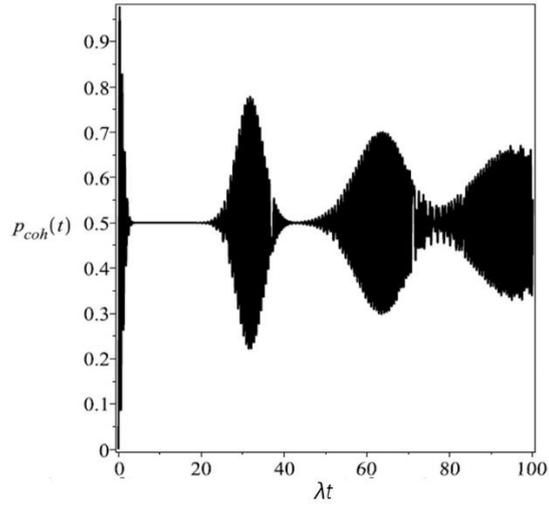

**FIG. 2:** Plot for the probability $p_{coh}(t)$ for the collapse and revival for the case of one-photon transition given in Eq. (21) for the coherent state of Fig. 1.

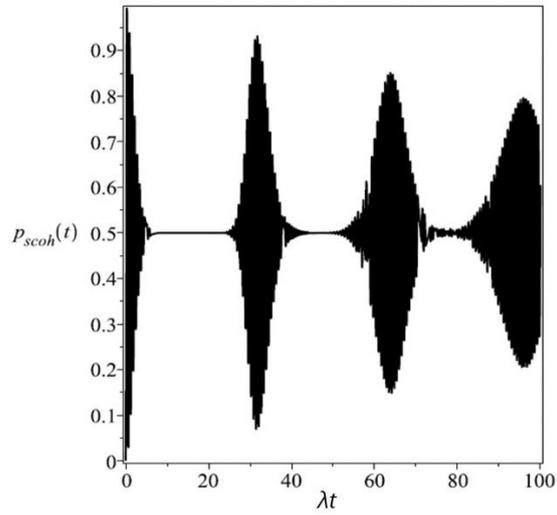

**FIG. 3:** Plot for the probability $p_{scoh}(t)$ for the collapse and revival for the case of one-photon transition given in Eq. (21) for the squeezed coherent state of Fig. 1.

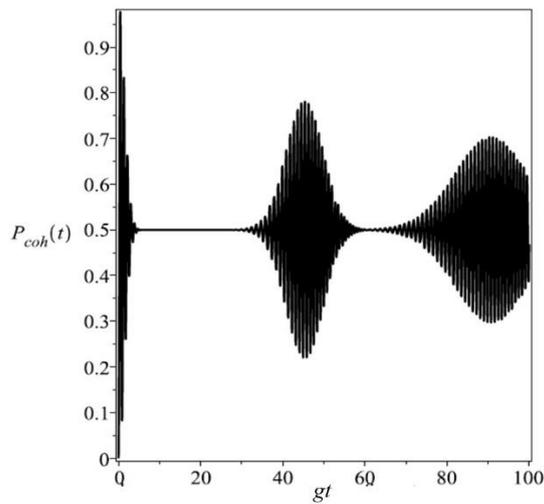

**FIG. 4:** Plot for the probability $P_{coh}(t)$ for the collapse and revival for the case of two-photon transition given in Eq. (22) for the coherent state of Fig. 1.

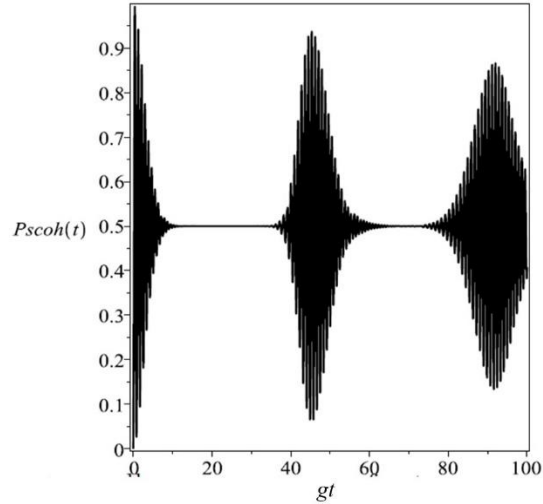

**FIG. 5:** Plot for the probability $P_{scoh}(t)$ for the collapse and revival for the case of two-photon transition given in Eq. (22) for the squeezed coherent state of Fig. 1.

where $\lambda$ is the coupling constant [15].

Fig. 1 shows the effect of squeezing for the probability of a coherent state (dash), viz., $|a_n|^2 = e^{-\bar{n}} \bar{n}^n / n!$, where $\bar{n}$ is the mean number of photons and $a_n = p(n)$ of Eq. (15) (solid). Note the effect on $p(n)$ of squeezing a coherent state with $\bar{n} = 100$ into a squeezed coherent state with $\bar{n} = 24.6$, in particular, the narrowing and increase of the peak at $n \approx \bar{n} = 24.6$.

Fig. 2 shows the probability $p_{coh}(t)$ for the coherent state and $p_{scoh}(t)$ in Fig. 3 for the squeezed coherent state for the collapse and revival. Note the narrowing and increase in the peaks in Fig. 3 as contrasted to those in Fig. 2, owing to the optimization of the squeezing of the coherent state.

The envelope of the oscillations collapse is a Gaussian with a $1/e$ time-width $\Omega_0 T_c = 2\sqrt{2}$ and the revival time is $\Omega_0 T_r = 4\pi\sqrt{\bar{n}}$, where $\Omega_0$ is the vacuum Rabi frequency measuring the atom-field coupling [16]. Now $\Omega_0 = 2\lambda$ and so $\lambda t_c = \sqrt{2}$ and $\lambda t_r = 2\pi\sqrt{\bar{n}}$ with values 1.41 and 31.2, respectively, which agree with the results in Fig. 2 and Fig. 3.

*b. Two-photon transition*

Three-level atoms interacting with a quantized electric field via the electric dipole Hamiltonian, whether the atom is in the $\Lambda$ or cascade configuration, can be reduced to an effective two-level atom undergoing two-photon transitions [17, 18]. Consider the atom initially in the excited state $|3\rangle$ while the radiation field is in a superposition of photon numbers with probability amplitude $a_n$, then the probability for being in the ground state $|1\rangle$ at time $t$ is

$$P(t) = \sum_{n=0}^{\infty} 2 \frac{(n+1)(n+2)}{(2n+3)^2} |a_n|^2 \left(1 - \cos\left(\sqrt{2n+3}\, gt\right)\right), \quad (22)$$

where the couplings $g_1 = g_2 \equiv g$ and the detuning $\Delta = 0$ in Ref. [17].

Fig. 4 and Fig. 5 shows the same as Fig. 2 and Fig. 3 albeit for the case of collapse and the revival governed by two-photon transitions as given by (22). Note again the strong effect of the optimization of the squeezing of the coherent state in sharpening the revival region for both one- and two-photon transitions.

Note also the displacement in time of the occurrence of both collapse and revival for both the coherent and squeezed coherent states compared to those for the one-photon transition. Now $\Omega_0 = \sqrt{2}g$ and so $gt_c = 2$ and $gt_r = 4\pi\sqrt{\bar{n}/2}$ with values 2.0 and 44.1, respectively, which agree with the results in Fig. (4) and Fig. (5).

## 7. Photon-number parity oscillations

The photon parity operator is defined by [16, 19]

$$\hat{\Pi} = (-1)^{\hat{n}} = \exp(i\pi\hat{n}), \tag{23}$$

where $\hat{n}$ is the photon number operator. The photon number parity is then

$$\Pi = \langle\hat{\Pi}\rangle = \langle\exp(i\pi\hat{n})\rangle = \sum_{n=0}^{\infty}(-1)^n p(n). \tag{24}$$

For instance, $\langle\hat{\Pi}\rangle = 1$ for squeezed states and $\langle\hat{\Pi}\rangle = \exp(-2\langle\hat{n}\rangle)$ for coherent states. Both of these results follow as limiting cases of the parity of the squeezed coherent state, viz., $\Pi_{scoh} = (|\alpha|, r) = \exp(-2|\alpha|^2)$ that follows from (15) and (16), where we suppose $\varphi = 2\theta$ and so the parity, expressed as a function of $|\alpha|$ and $r$, is actually independent of $r$. However, $|\alpha|^2$ is not equal to the mean number of photons $\bar{n}$, instead, it is a function of $\bar{n}$ and the squeezing parameter $r$ as given in (18). Therefore, the parity of the squeezed coherent state is a function of both $\bar{n}$ and $r$.

*a. One-photon transition*

The photon number parity for the one-photon transition is given by

$$\Pi_1(t) = \langle\hat{\Pi}_1(t)\rangle = \frac{1}{2}\sum_{n=0}^{\infty}(-1)^n |a_n|^2 \left[1 - \cos(2\lambda\sqrt{n+1}\,t)\right]. \tag{25}$$

The result for the coherent state is shown in Fig. 6 and for the squeezed coherent state in Fig. 7, where the parameters are as in Fig. 1, viz., $\langle\hat{n}\rangle = 24.6, |\alpha| = 10,$ and $r = 0.7136$. We note again the narrowing and the increase in the peak heights in the revival regions.

It is interesting that close to half the revival time, $t_e = t_r/2$, the average photon number parity for the coherent state is rapidly oscillating, and that this is also the time at which the field and atomic states are separable [19]. For the coherent state with parameters given in Fig. 1, $\lambda t_e = \pi\sqrt{\bar{n}} = 15.6$, which is quite consistent with the results show in Fig. 6. This is also true for the squeezed coherent state shown in Fig. 7.

b.  *Two-photon transition*

The photon number parity for the two-photon transition is given by

$$\Pi_2(t) = \langle \hat{\Pi}_2(t) \rangle = \sum_{n=0}^{\infty} 2(-1)^n \frac{(n+1)(n+2)}{(2n+3)^2} |a_n|^2 \left[1 - \cos(\sqrt{2n+3}\, gt)\right]. \qquad (26)$$

The result for the coherent state is shown in Fig. 8 and Fig. 9 corresponds to the squeezed coherent state, where the parameters are as in Fig. 1, viz., $\langle \hat{n} \rangle = 24.6$, $|\alpha| = 10$, and $r = 0.7136$. We note again the narrowing and the increase in the peak heights in the revival regions. The Rabi frequency oscillations of the photon number parity in this case occurs first at $gt_e = 2\pi\sqrt{\bar{n}/2} = 22$, which agrees with the results in both Fig. 8 and Fig. 9.

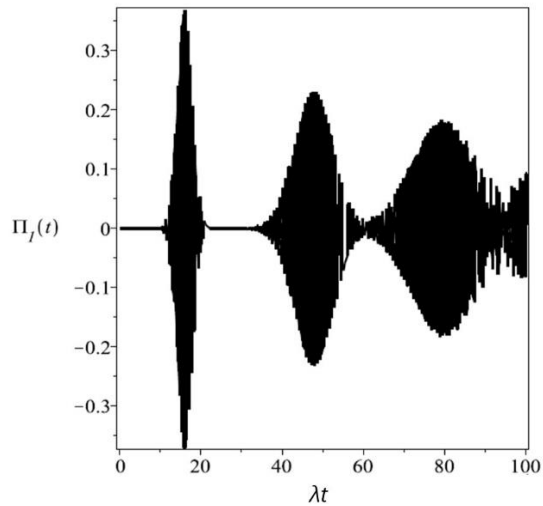

**FIG. 6:** Plot of the expectation value of the field parity operator given in Eq. (25) for the coherent state for the one-photon transition as for Fig. 1.

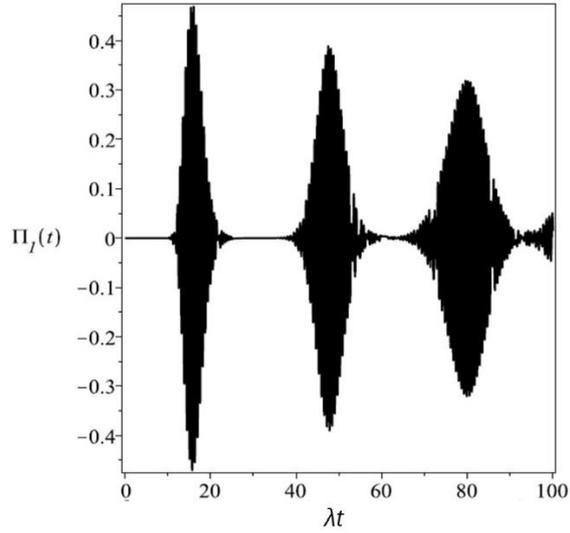

**FIG. 7:** Plot of the expectation value of the field parity operator given in Eq. (25) for the squeezed coherent state for the one-photon transition as for Fig. 1.

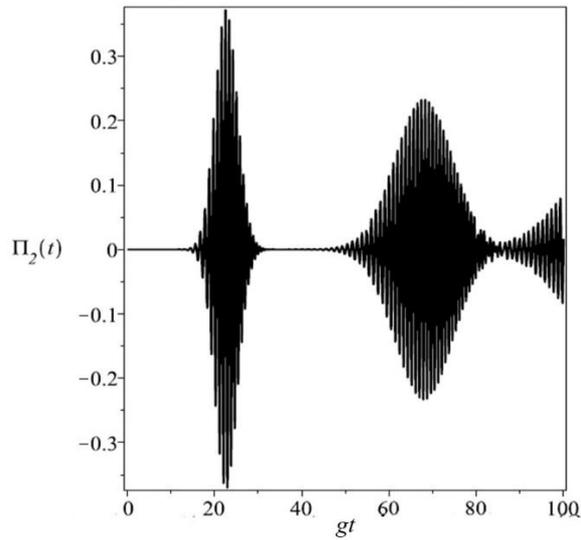

**FIG. 8:** Plot of the expectation value of the field parity operator given in Eq. (26) for the coherent state for the two-photon transition as for Fig. 1.

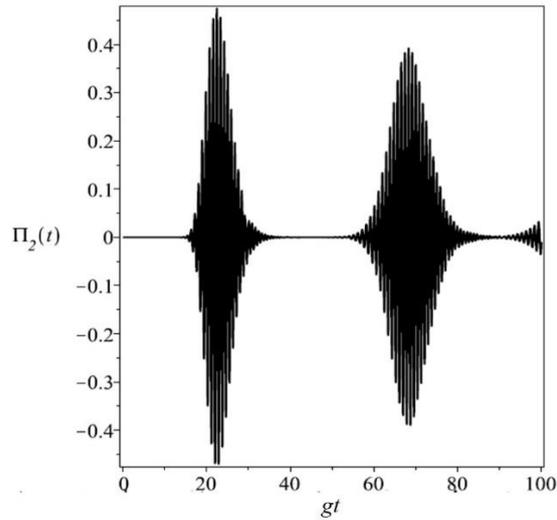

**FIG. 9:** Plot of the expectation value of the field parity operator given in Eq. (26) for the squeezed coherent state for the two-photon transition as for Fig. 1.

## 8. Summary and discussion

We use the Bogoliubov-Valatin transformation to introduce the notion of quasiparticles describing the radiation field. An $n-$quasiparticle state is equal to a squeezed $n-$photon state and so the quasiparticle states span the whole Hilbert space of photons. Accordingly, the quasiparticle coherent state is equal to the squeezed coherent state of photons. In order to sharpen the region where revival occurs, the ratio of photon variance to photon number is minimized, which gives, for a given value of $|\alpha|$, both the magnitude of squeezing, $r$, and the mean photon number $\bar{n}$. The effect of squeezing the coherent state is to reduce considerably the mean photon number in the coherent state to a much lower mean photon number in the squeezed coherent state. In the numerical case considered, the reduction is by $75\%$, from a mean number of photons of $100$ to $25$. Experimental revival observations have so far been limited to small photon numbers since experiments face formidable challenges. Therefore, it is hoped that squeezing photon states may reduce the number of photons from mesoscopic to microscopic and so make experimental observations possible but also amenable to simpler numerical simulation with the aid of the Jaynes-Cummings model.